\documentclass[pdflatex,sn-mathphys-ay]{sn-jnl}

\usepackage{graphicx}%
\usepackage{multirow}%
\usepackage{amsmath,amssymb,amsfonts}%
\usepackage{amsthm}%
\usepackage{mathrsfs}%
\usepackage[title]{appendix}%
\usepackage{xcolor}%
\usepackage{textcomp}%
\usepackage{manyfoot}%
\usepackage{booktabs}%
\usepackage{algorithm}%
\usepackage{algorithmicx}%
\usepackage{algpseudocode}%
\usepackage{listings}%

\usepackage{mathtools}
\usepackage{diagbox}
\usepackage{makecell}
\usepackage{adjustbox}
\usepackage[normalem]{ulem}
\newcommand{\celltab}[1]{%
  \begin{tabular}{@{}r r r@{}}
  \multicolumn{1}{c}{$k$} & \multicolumn{1}{c}{$L_k$} & \multicolumn{1}{c}{\#} \\
  #1\\
  \end{tabular}%
}

\usepackage{soul}

\theoremstyle{thmstyleone}%
\theoremstyle{thmstyletwo}%

\theoremstyle{thmstylethree}%

\begin{document}

\let\vec\boldsymbol
\def\E{\mathrm{E}}
\def\inc{I}
\def\mstar{M_{\bullet}}
\def\real{\operatorname{Re}}
\def\imag{\operatorname{Im}}
\def\L{\mathrm{L}}

\def\vece{\vec{e}}
\def\vech{\vec{h}}
\def\vecrho{\vec{\rho}}
\def\vecz{\vec{z}}
\def\veczeta{\vec{\zeta}}

\def\vecptilde{\vec{\tilde{p}}}
\def\vecrhotilde{\vec{\tilde{\rho}}}

\newcommand{\vechat}[1]{\vec{\hat{#1}}}
\newcommand{\rot}[1]{R_{\vec{\hat{#1}}}}


\newcommand{\araa}{Annual Review of Astronomy and Astrophysics}
\newcommand{\apj}{Astrophysical Journal}
\newcommand{\apjs}{Astrophysical Journal, Supplement}
\newcommand{\apjl}{Astrophysical Journal, Letters}
\newcommand{\aap}{Astronomy and Astrophysics}
\newcommand{\icarus}{Icarus}
\newcommand{\mnras}{Monthly Notices of the RAS}
\newcommand{\nat}{Nature}
\newcommand{\physrep}{Physics Reports}
\newcommand{\na}{New Astronomy}
\newcommand{\aj}{Astronomical Journal}
\newcommand{\pasp}{Publications of the ASP}
\newcommand{\actaa}{Acta Astronomica}
\newcommand{\pasj}{Publications of the Astronomical Society of Japan}
\newcommand{\zap}{Zeitschrift fuer Astrophysik}

\title[Article Title]{Laplace-Legendre expansion of the planar planetary Hamiltonian}


\author*[1]{\fnm{Aya} \sur{Alnajjarine}}\email{aya.alnajjarine@obspm.fr}
\author*[1]{\fnm{Jacques} \sur{Laskar}}\email{jacques.laskar@obspm.fr}
\author[2,1]{\fnm{Federico} \sur{Mogavero}}\email{mogavero@math.unipd.it}

\affil[1]{\orgdiv{LTE, CNRS, Observatoire de Paris}, \orgname{Universit\'{e} PSL, Sorbonne Universit\'{e}}, \orgaddress{\street{77 Avenue Denfert-Rochereau}, \postcode{75014} \city{Paris}, \country{France}}}
\affil[2]{\orgdiv{Dipartimento di Matematica “Tullio Levi-Civita”}, \orgname{Università degli Studi di Padova}, \orgaddress{\street{Via Trieste 63}, \postcode{35121} \city{Padova}, \country{Italy}}}



\abstract{We explore a hybrid expansion of the disturbing function in planetary dynamics that combines elements of the classical Laplace and Legendre developments. This formulation retains the structure of the Laplace expansion, but expresses the inverse of the mutual distance as a series whose terms keep an exact dependence on both the eccentricity and the semi-major axis ratio. We use it to construct the first-order secular Hamiltonian of the planar 3-body problem, relevant for modeling the long-term evolution of planetary systems. We assess the convergence of the new expansion numerically and compare it with that of the Laplace and Legendre series across a range of orbital configurations. The results show that the new expansion provides consistent performance across diverse dynamical regimes, bridging the domains of applicability of the two classical approaches.}




\maketitle

\section{Introduction}
\label{sec:Intro}
In planetary systems, the disturbing function plays a central role in understanding the long-term dynamical evolution of planetary orbits. It encodes the gravitational interactions between planets and provides the foundation for studying both secular and resonant dynamics. Expanding the disturbing function as a power series in the orbital elements has long been a key focus in celestial mechanics and fundamental to constructing perturbative models of planetary motion. In the context of the 3-body problem, two classical approaches are commonly used for this purpose.

The first approach expands the disturbing function as a power series in eccentricity and inclination, with coefficients that are exact in semi-major axes ratio and are expressed in terms of Laplace coefficients. This formulation, developed by Laplace \citeyearpar{Laplace1773} and Lagrange \citeyearpar{Lagrange1781,Lagrange1782}, forms the basis of what is now known as the Laplace–Lagrange secular theory \citep[see][for a detailed historical account]{laskar2013}. This theory has been a cornerstone for studying the long-term evolution of planetary systems, particularly in the solar system, where orbits are nearly circular and coplanar. Peirce \citeyearpar{Peirce1849} and Le Verrier \citeyearpar{LeVerrier1855} extended the expansion to higher degrees in eccentricities and inclinations, aiming for a more accurate description of the solar system’s evolution, followed by further similar developments \citep{Boquet1889,Newcomb1895,BrouwerClemence1961,Murray1999}. This expansion was later reformulated within a Hamiltonian framework using heliocentric canonical variables introduced by Poincaré \citep{Poincare1896,Laskar1991,Laskar1995}. 

The second approach develops the disturbing function in powers of the semi-major axes ratio of the two planets, using Legendre polynomials. This expansion, known as Legendre expansion, has been widely applied to a variety of triple systems with hierarchical configurations  \citep{Hansen1855,hill1875,Tisserand1894,Kaula1962,Kozai1962,Lidov1962,brumberg1967,Ford2000,Lee2003,Laskar2010,Naoz2011}, and is applicable to configurations with arbitrary eccentricities and inclinations.

Each expansion offers advantages in different dynamical regimes. Laplace expansion is well suited to compact, nearly circular and coplanar orbits, as in the solar system, but has limitations when applied to configurations with higher eccentricities and inclinations, such as those observed in certain exoplanetary systems. In contrast, Legendre expansion is applicable for objects with arbitrary eccentricities and inclinations, but converges slowly when the system is not strongly hierarchical, as in the case of the inner solar system \citep{Boue2012}. These complementary limitations naturally raise the question of whether an alternative formulation can be constructed to combine both approaches, and whether such a formulation could leverage the advantages of both. \citet{Ellis2000} presented a mixed expansion of the disturbing function “\textit{which retains the advantages of \citet{Kaula1962}’s approach but expressed in terms of Laplace coefficients and their derivatives}". The final form of such an expansion is given in their Eq.~(16), but its detailed derivation was deferred to a subsequent paper that does not appear to have been published \citep[see also][]{Mardling2013}. The new development was only used as a basis of a “\textit{for producing expansions of the planetary disturbing function to any order in the eccentricities and inclinations}", and no study of its convergence was performed. 

In this work, we explore an alternative development of the disturbing function, which we refer to as the Laplace–Legendre expansion. It retains the classical structure of the Laplace expansion, developing the inverse of the mutual distance as a Taylor series around the circular, coplanar problem. The key difference is that, at each order, the terms of the new expansion are exact in both eccentricity and the semi-major axis ratio. We present the new formulation for the planar 3-body problem, with a specific focus on its secular part, which is relevant for modeling the long-term orbital evolution of planetary systems. 

We begin by outlining the Hamiltonian formalism for the planetary 3-body problem and reviewing the two classical approaches for expanding the direct part of the disturbing function (Sect.~\ref{sec:threebodyHamiltonian}). We then introduce the Laplace–Legendre expansion at zeroth, first, and arbitrary order (Sect.~\ref{sec:LL_formulation}). Finally, we assess its convergence properties, comparing its accuracy with that of the classical Laplace and Legendre expansions across a range of eccentricities and semi-major axis ratios (Sect.~\ref{sec:numerical_convergence}).

\section{Hamiltonian formulation and expansion of the disturbing function}
\label{sec:threebodyHamiltonian}
We consider a 3-body system consisting of a central star of mass $m_0$ and two planets of masses $m_1$ and $m_2$, with $m_i \ll m_0$. Using canonical heliocentric variables \citep{Poincare1896,Laskar1991}, the Hamiltonian governing the Newtonian dynamics reads
\begin{equation}
\label{eq:Hamiltonian}
H = \underbrace{\sum_{i=1}^2 \left( \frac{\Vert \vec{\tilde{r}}_i \Vert^2}{2 \beta_i} 
- \frac{\mu_i \beta_i}{\Vert \vec{r}_i \Vert} \right)}_{H_0}
+ \underbrace{\vphantom{\sum_{i=1}^2} \frac{\vec{\tilde{r}}_1 \cdot \vec{\tilde{r}}_2}{m_0} }_{T_1} \
\underbrace{\vphantom{\sum_{i=1}^2} -\frac{G m_1 m_2}{\Vert \vec{r}_1 - \vec{r}_2 \Vert}}_{U_1},
\end{equation}
where $\vec{r}_i$ are the heliocentric position vectors of the planets, $\vec{\tilde{r}}_i$ are their conjugate barycentric momenta, $G$ the gravitational constant, $\mu_i = G(m_0 + m_i)$, and $\beta_i = m_0 m_i/(m_0 + m_i)$. The indices $i = 1, 2$ refer to the inner and outer planets, respectively.
The Keplerian part of the Hamiltonian $H_0$ describes decoupled Keplerian motions, each governed by the classical orbital elements: semi-major axis $a$, eccentricity $e$, inclination $I$, mean anomaly $M$, argument of pericenter $\omega$, and longitude of ascending node $\Omega$. In terms of these elements, the Keplerian part takes the form $H_0 = -\sum_{i=1}^{2} \mu_i \beta_i / 2 a_i$.

The perturbation to this integrable part consists of two terms: an indirect part $T_1$, arising from the non-inertial nature of the heliocentric frame, and a direct part $U_1$, describing the mutual gravitational interaction between the planets. The indirect term $T_1$ averages to zero over the fast orbital angles $M_i$ and thus does not contribute to the averaged Hamiltonian in secular theories limited to first-order in
planetary masses. We therefore focus in this work on the expansion of the direct part $U_1$, which encapsulates the main complexity of the perturbing function and governs the secular dynamics of planetary systems. We focus primarily on the planar problem and begin by summarizing the two classical methods used to expand this term.

\subsection{Legendre expansion of the direct part}
\label{sec:Legendre_Expansion}
We denote the mutual distance between the planets as $\Delta = \Vert \vec{r}_1 - \vec{r}_2 \Vert$. Its inverse can be expanded using Legendre polynomials as
\begin{equation}
\label{eq:directpart}
\frac{a_2}{\Delta}=\frac{a_2}{r_2} \sum_{n \in \mathbb N} \left(\frac{r_1}{r_2}\right)^n P_n (\cos S),
\end{equation}
where $S$ is the angle between the position vectors $\vec{r}_1$ and $\vec{r}_2$, and $P_n$ denotes the Legendre polynomial of degree $n$. Following the formalism of \citep{Laskar2010}, this expression can be transformed into a Fourier series in the mean anomalies $M_i$ and written as a power series in the semi-major axis ratio $\alpha = a_1 / a_2$,
\begin{equation}
\label{eq:directpart_Legendre}
    U_1 = -\frac{Gm_1m_2}{a_2} \ \sum_{n \in\mathbb{N}} \mathcal{F}_n\alpha^n \quad ; \quad \mathcal{F}_n=\sum_{k_1,k_2 \in \mathbb Z} \mathcal{F}_n^{(k_1,k_2)}\E^{\iota (k_1M_1+k_2M_2)},
\end{equation}
where the indices $k_i$  label the harmonics of the mean anomalies $M_i$. 
In all this work, we assume that we are in a domain $\cal D$ excluding  collisions, that is $ |r_1/r_2| < \rho_0 < 1 $. In this case $1/\Delta$ is analytical on $\cal D$ and its Fourier expansion with double arguments $(M_1, M_2)$ (Eq. \ref{eq:directpart_Legendre}) is absolutely and uniformly convergent with coefficients that decrease exponentially with the order ($|k_1|, |k_2|$).
Throughout this paper, $\iota$ represents the imaginary unit and $\E$ the exponential operator.
In the planar case, the coefficients $\mathcal{F}_n^{(k_1,k_2)}$ depend on the orbital eccentricities $e_i$ and the arguments of pericenter  $\omega_i$. They take the form
\begin{equation}
\label{eq:Legendre_coeff}
\mathcal{F}_n^{(k_1,k_2)} = \sum_{q=0}^{n} f_{n,q} \, X_{k_1}^{n,2q-n}(e_1) \, X_{k_2}^{-(n+1),n-2q}(e_2) \, \E^{\iota (2q - n)(\omega_1 - \omega_2)},
\end{equation}
where $f_{n,q}$ are constant coefficients given by
\begin{equation}
\label{eq:fnq_coeff}
    f_{n,q}= \frac{(2q)!(2n-2q)!}{2^{2n}(q!)^2((n-q)!)^2}, 
\end{equation}
for $0\leq q \leq n$. The functions $X_k^{n,m}(e)$ are Hansen coefficients, defined for $n,m\in \mathbb Z$ as 
\begin{equation}
\label{eq:hansen_coeff}
\left(\frac{r}{a}\right)^n \E^{\iota m\nu} = \sum_{k\in \mathbb{Z}}X_k^{n,m}(e) \, \E^{\iota kM},
\end{equation}
where $\nu$ is the true anomaly \citep{Hansen1855,Tisserand1894}. They are polynomials in the eccentricity and are related to each other via recurrence relations \citep[e.g.,][and references therein]{Laskar2010}.
Equations~\eqref{eq:directpart_Legendre},~\eqref{eq:Legendre_coeff} and~\eqref{eq:fnq_coeff} provide an explicit algorithm for computing the direct part of the disturbing function $U_1$, and in particular for extracting its secular contribution up to any order $N$ in $\alpha$,
\begin{equation}
\label{eq:Legendre_secular}
\langle U_1 \rangle \ = -\frac{Gm_1m_2}{a_2} \sum_{n=0}^{N} \mathcal{F}_n^{(0,0)} \alpha^n.
\end{equation}

\subsection{Laplace expansion of the direct part}
\label{sec:Laplace_expansion}
For systems with low eccentricities and inclinations, the inverse of the mutual distance between planets can be expanded around the circular planar problem. Following the formalism of \citep{Laskar1991,Laskar1995}, it is expressed as 
\begin{equation}
\label{eq:distance_Laplace}
\frac{a_2}{\Delta} = \frac{a_2}{r_2} (A+V)^{-1/2} = \frac{a_2}{r_2}\sum_{k \in \mathbb{N}} (-1)^k \frac{(2k)!}{4^k (k!)^2} V^k A^{-k-1/2}, 
\end{equation}
where $A$ and $V$ are defined as 
\begin{equation}
\label{eq:Laplace_A_V}
\begin{aligned}
A&= 1-2\alpha\cos{(\lambda_1-\lambda_2)}+\alpha^2,\\
V &= 2\alpha \left( \cos{(\lambda_1-\lambda_2)} - \frac{\rho}{\alpha} \cos S \right) + \alpha^2 \left( \left( \frac{\rho}{\alpha}\right)^2 - 1\right).\\
\end{aligned}
\end{equation}
Here, $\rho=r_1/r_2$, $\lambda_i = M_i + \omega_i + \Omega_i$ are the planetary mean longitudes and $S$ is the angle between the heliocentric position vectors $\vec{r}_1$ and $\vec{r}_2$. The term $A$  corresponds to the square of the mutual distance in the circular planar case and admits the Fourier expansion
\begin{equation}
\label{eq:Laplace_coef}
A^{-s} = \frac{1}{2}\sum_{l \in \mathbb{Z}} b_s^{(l)}(\alpha) \, \E^{\iota l (\lambda_1 - \lambda_2)},
\end{equation}
where $b_s^{(l)}(\alpha)$ are the classical Laplace coefficients \citep{Laplace1878}. 
As long as $0<\alpha< \alpha_0<1$, $A$ is bounded from zero and the Fourier series \eqref{eq:Laplace_coef} converges absolutely and uniformly with Laplace coefficients that decrease exponentially with the order $l$. 
The Taylor expansion of $(1+x)^{-1/2}$ about $x=0$ converges for $|x|<1$  and has radius of convergence $1$. 
The series \eqref{eq:distance_Laplace} is thus convergent for $|V| < |A|$.

We have $A > (1-\alpha)^2$. On the other hand, $V$ collects the terms involving eccentricities and inclinations, starting at degree 1 in eccentricity, degree 2 in inclination, and has $\alpha$ in factor. Thus, for sufficiently small $\alpha$, eccentricities and inclinations, $|V| < |A|$ is satisfied and the series \eqref{eq:distance_Laplace} is convergent.

As in \citep{Laskar1991,Laskar1995}, we express this expansion in terms of Poincaré variables, which, in the planar case, take the form
\begin{equation}
\Lambda_i = \beta_i \sqrt{\mu_i a_i}, \qquad x_i = \sqrt{\Lambda_i} \sqrt{1 - \sqrt{1 - e_i^2}} \, \E^{\iota \omega_i}. 
\end{equation}
The pairs $(\Lambda_i, \lambda_i)$ and $(x_i, -\iota \bar{x}_i)$ are canonical conjugate momentum-coordinate variables, with the overbar denoting complex conjugation.

In these variables, the direct part of the perturbation $U_1$ can be written as a Fourier series in the mean longitudes, whose coefficients are polynomial series depending on the remaining Poincaré variables  \citep{Laskar1990b,Laskar1991,Laskar1995}, 
\begin{equation}
\label{eq:Laplace_planar}
\begin{aligned}
&U_1 = -\frac{Gm_1m_2}{a_2} \sum_{l_1, l_2 \in \mathbb{Z}} C_{l_1,l_2} \, \E^{\iota (l_1 \lambda_1 + l_2 \lambda_2)}, \\
&C_{l_1,l_2} = \sum_{\vec{n} \in \mathbb{N}^4} \Upsilon_{\vec{n}}(\alpha) \, x_1^{n_1} \bar{x}_1^{\bar{n}_1} x_2^{n_2} \bar{x}_2^{\bar{n}_2},
\end{aligned}
\end{equation}
where $\vec n = (n_1,\bar{n}_1,n_2,\bar{n}_2)$ and the coefficients $\Upsilon_{\vec{n}}(\alpha)$ can be expressed analytically in terms of Laplace coefficients \citep[and references therein]{Laskar1995}. This formulation provides a systematic expansion of $U_1$ in powers of the eccentricities, with explicit expressions for its secular component $ C_{0,0}$ at any desired degree.

\section{Laplace-Legendre expansion}
\label{sec:LL_formulation}
Laplace expansion given in Eq.~\eqref{eq:distance_Laplace} is typically developed by expanding each term in powers of eccentricity and inclination. In this section, we introduce an alternative formulation, which we call the Laplace-Legendre expansion as it combines elements from both the Laplace and Legendre approaches. It has the same structure as the Laplace expansion, but at each order, the retained terms keep their full dependence in both eccentricity and $\alpha$ and are expressed using both Laplace and Hansen coefficients. We develop this formulation here for the planar case.

We recall that in the limit of small eccentricities, $V$ can be approximated as
\begin{equation}
\label{eq:V_small_e}
\begin{aligned}
V = & \, 2\alpha \biggl( 
  \frac{3}{2} e_1 \cos(\omega_1 - \lambda_2) - \frac{3}{2} e_2 \cos(\lambda_1 - 2\lambda_2 + \omega_2)\\
&- \frac{1}{2} e_1 \cos(2\lambda_1 - \omega_1 - \lambda_2) + \frac{1}{2} e_2 \cos(\lambda_1 - \omega_2)
\biggr) \\
& + \alpha^2 \biggl(  -2 e_1 \cos(\lambda_1 - \omega_1) + 2 e_2 \cos(\lambda_2 - \omega_2) \biggr)
+ \mathcal{O}(e^2).
\end{aligned}
\end{equation}
So $V$ is a small quantity when either $\alpha$ or $e$ are small. Our idea is, therefore, to retain the exact dependence of $V$ on $e$ and $\alpha$ at each order, avoiding the expansion in $e$ carried on in the classical Laplace approach. The resulting expansion is expected to converge better than the classical developments in regimes where both $e$ and $\alpha$ are small, or in intermediate regimes where neither parameter is particularly small but their product remains small. These regimes define the natural domain of interest of the present formulation, which is explored quantitatively in Sect.~\ref{sec:numerical_convergence}.Moreover, exact dependence on $e$ and $\alpha$ at each order should allow us to span the two regimes $e\to0$ and $\alpha\to0$ in a single formulation. In the following, we construct the Laplace–Legendre expansion at zeroth, first, and arbitrary order, focusing in each case on the expression of the secular direct part of the disturbing function.

\subsection{Laplace-Legendre expansion of order 0}
\label{sec:LL_order0}
We define the Laplace-Legendre expansion of order 0 as the term
\begin{equation}
\label{eq:LL_order0}
\frac{a_2}{\Delta}  = \frac{a_2}{r_2} A^{-1/2},
\end{equation}
which corresponds to the first term of the Laplace expansion and is of degree zero in eccentricities. We denote this term by $\mathcal{A}_0$ and aim to express it in a form that is exact in both $e$ and $\alpha$, using Laplace and Hansen coefficients.  As in classical secular theories, we aim to express $\mathcal{A}_0$ as a Fourier series in the mean longitudes $\lambda_i$ in order to extract its Fourier components $\mathcal A_0^{(k_1,k_2)}$, i.e. 
\begin{equation}
\mathcal A_0 = \sum_{k_1,k_2 \in \mathbb{Z}} \mathcal{A}_0^{(k_1,k_2)} \E^{\iota(k_1\lambda_1 + k_2\lambda_2)}.
\end{equation}
We begin by writing
\begin{equation}
    \mathcal{A}_0= \frac{1}{2}\frac{a_2}{r_2}\sum_{l\in \mathbb{Z}} b_{1/2}^{(l)}(\alpha) \E^{\iota l(\lambda_1-\lambda_2)}.
\end{equation}
Using the Hansen coefficients defined in Eq.~\eqref{eq:hansen_coeff}, the ratio $a_2/r_2$ can be expanded as
\begin{equation}
\label{eq:a2_r2}
\left(\frac{a_2}{r_2}\right)= \sum_{k_2\in \mathbb{Z}}X_{k_2}^{-1,0}(e_2) \, \E^{\iota k_2 M_2}.
\end{equation}
Substituting this into the expression of $\mathcal{A}_0$, we obtain
\begin{equation}
\mathcal{A}_0 = \frac{1}{2} \sum_{l, k_2 \in \mathbb{Z}} b_{1/2}^{(l)}(\alpha) X_{k_2}^{-1,0}(e_2) \E^{\iota l(\lambda_1-\lambda_2)} \E^{\iota k_2(\lambda_2 - \omega_2)}.
\end{equation}
The secular component corresponds to the $(0,0)$ harmonic in the mean longitudes, i.e., $l=0$ and $k_2= 0$. Knowing that $X_0^{-1,0}(e_2) = 1$, the secular component reduces to
\begin{equation}
\label{eq:A0_Laplace}
\mathcal{A}_0^{(0,0)} = \frac{1}{2} b_{1/2}^{(0)}(\alpha),
\end{equation}
recovering the classical Laplace expression at degree zero in eccentricities.
Accordingly, the secular contribution of the direct part $U_1$ is given by
\begin{equation}
\langle U_1\rangle = -\frac{Gm_1m_2}{a_2}  \mathcal{A}_0^{(0,0)}.
\end{equation}

\subsection{Laplace-Legendre expansion of order 1}
\label{sec:LL_order1}
We now introduce the Laplace–Legendre expansion of order 1, defined by 
\begin{equation}
\frac{a_2}{\Delta} = \mathcal{A}_0 + \mathcal{A}_1 ,
\end{equation}
where $\mathcal{A}_1$ corresponds to the second term in the Laplace expansion, that is,
\begin{equation}
\mathcal{A}_1 = -\frac{1}{2} \frac{a_2}{r_2} V A^{-3/2}.
\end{equation}
As before, our goal is to express $\mathcal{A}_1$ in a form that is exact in $e$ and $\alpha$. Recall that $V$ takes the form
\begin{equation}
\label{eq:Laplace_Vexp2}
V = 2\alpha \left( \cos{(\lambda_1-\lambda_2)} - \frac{\rho}{\alpha} \cos S \right) + \alpha^2 \left( \left( \frac{\rho}{\alpha}\right)^2 - 1\right),
\end{equation}
where $S = (\nu_1 + \omega_1) - (\nu_2 + \omega_2)$ in the planar case, and $\rho / \alpha = (r_1 / a_1)(a_2 / r_2)$.
To proceed, we express $A^{-3/2}$ using Laplace coefficients as
\begin{equation}
A^{-3/2} = \frac{1}{2} \sum_{l \in \mathbb{Z}} b_{3/2}^{(l)}(\alpha)  \E^{\iota l (\lambda_1 - \lambda_2)}.
\end{equation}
We next expand $\dfrac{a_2}{r_2} V$ as a Fourier series in the mean longitudes $\lambda_i$, using Hansen coefficients (see Eq.~\ref{eq:hansen_coeff}). This yields
\begin{equation}
\label{eq:V_lambda}
\begin{split}
\frac{a_2}{r_2} V =& \, \alpha \sum_{k_2 \in \mathbb{Z}} X_{k_2}^{-1,0} \, \E^{\iota k_2(\lambda_2 - \omega_2)} \left( 2 \cos(\lambda_1 - \lambda_2) - \alpha \right) \\
&-\alpha \sum_{k_1, k_2 \in \mathbb{Z}} \E^{\iota[k_1(\lambda_1 - \omega_1) + k_2(\lambda_2 - \omega_2)]} \\
&\quad \times \left( X_{k_1}^{1,1} X_{k_2}^{-2,-1} \E^{\iota(\omega_1 - \omega_2)} + X_{k_1}^{1,-1} X_{k_2}^{-2,1} \E^{-\iota(\omega_1 - \omega_2)} - \alpha X_{k_1}^{2,0} X_{k_2}^{-3,0} \right).
\end{split}
\end{equation} 
Substituting this expression along with the expansion of $A^{-3/2}$ into $\mathcal{A}_1$, we obtain
\begin{equation}
\label{eq:A1_Laplace}
\begin{aligned}
\mathcal{A}_1 = & -\frac{\alpha}{4} \sum_{l, k_2 \in \mathbb{Z}} b_{3/2}^{(l)}(\alpha) X_{k_2}^{-1,0} \, \E^{\iota k_2(\lambda_2 - \omega_2)} \\
& \qquad \times \left( \E^{\iota(l+1)(\lambda_1 - \lambda_2)} + \E^{\iota(l-1)(\lambda_1 - \lambda_2)} - \alpha \, \E^{\iota l(\lambda_1 - \lambda_2)} \right) \\[1ex]
& + \frac{\alpha}{4} \sum_{l, k_1, k_2 \in \mathbb{Z}} b_{3/2}^{(l)}(\alpha) \, \E^{\iota[(k_1 + l)\lambda_1 + (k_2 - l)\lambda_2]} \, \E^{-\iota(k_1 \omega_1 + k_2 \omega_2)} \\
& \qquad \times \left( X_{k_1}^{1,1} X_{k_2}^{-2,-1} \E^{\iota(\omega_1 - \omega_2)} + X_{k_1}^{1,-1} X_{k_2}^{-2,1} \E^{-\iota(\omega_1 - \omega_2)} - \alpha X_{k_1}^{2,0} X_{k_2}^{-3,0} \right).
\end{aligned}
\end{equation} 
For convenience, we omit the explicit dependence on eccentricities $e_1$ and $e_2$ in the Hansen coefficients $X_{k_1}^{n,m}(e_1)$ and $X_{k_2}^{n,m}(e_2)$, unless needed for clarity. The indices $k_1$ and $k_2$ refer to the inner and outer planet, respectively. 

To isolate the secular part of $\mathcal{A}_1$, we retain only terms with zero frequency in the mean longitudes. In the first sum, this corresponds to $k_2 = 0$ and $l \in \{ -1, 0, 1 \}$. In the second sum, secular terms correspond to $k_1 = -l$ and $k_2 = l$. This yields
\begin{equation}
\label{eq:A1_Laplace_secular}
\begin{aligned}
\mathcal{A}_{1}^{(0,0)} =& -\frac{\alpha}{4} \left( b_{3/2}^{(-1)}(\alpha) + b_{3/2}^{(1)}(\alpha) - \alpha b_{3/2}^{(0)}(\alpha) \right) \\[1ex]
& + \frac{\alpha}{4} \sum_{l \in \mathbb{Z}} b_{3/2}^{(l)}(\alpha) \biggl[ 
   - \alpha \, X_{-l}^{2,0}(e_1) \, X_l^{-3,0}(e_2) \, \E^{\iota l(\omega_1 - \omega_2)} \\
& \hspace{7.5em}
+ X_{-l}^{1,1}(e_1) \, X_l^{-2,-1}(e_2) \, \E^{\iota(l+1)(\omega_1 - \omega_2)} \\
& \hspace{7.5em}
+ X_{-l}^{1,-1}(e_1) \, X_l^{-2,1}(e_2) \, \E^{\iota(l-1)(\omega_1 - \omega_2)} \biggr].
\end{aligned}
\end{equation} 
Accordingly, the secular contribution of $U_1$ reads
\begin{equation}
\langle U_1 \rangle= -\frac{Gm_1m_2}{a_2} \left(\mathcal{A}_0^{(0,0)} + \mathcal{A}_1^{(0,0)}  \right).
\end{equation}

\subsection{Generalization to arbitrary order}
\label{sec:LL_general_order}
The formalism presented at orders $0$ and $1$ for this mixed expansion can be systematically extended to arbitrary order. At each order $k$, we express $a_2/\Delta$ as a sum of the first $k+1$ terms of the classical Laplace expansion, with each term reformulated in a way that is exact in both $e$ and $\alpha$. Specifically, we write
\begin{equation}
\frac{a_2}{\Delta} = \sum_{j=0}^{k} \mathcal{A}_j,
\end{equation}
where each term $\mathcal{A}_j$ takes the form
\begin{equation}
\mathcal{A}_j = \frac{a_2}{r_2}  C_j V^{j} A^{-j - 1/2}, \qquad C_j = (-1)^{j} \frac{(2j)!}{4^{j} j!^2}.
\end{equation}
To extract the secular component from each term, we follow the procedure outlined for orders $0$ and $1$. The term $A^{-j - 1/2}$ is expanded as a Fourier series in the angle $\lambda_1 - \lambda_2$ using Laplace coefficients:
\begin{equation}
A^{-s} = \frac{1}{2}\sum_{l \in \mathbb{Z}} b_s^{(l)}(\alpha) \, \E^{\iota l (\lambda_1 - \lambda_2)}.
\end{equation}
The expression $V^j$ consists of a finite sum of products of trigonometric terms involving the angles $(\lambda_i, \omega_i, \nu_i)$ and powers of the ratios $r_i/a_i$. The product $(a_2/r_2)  V^{j}$ can thus be expressed as a series in the mean anomalies $M_i$, using Hansen coefficients defined in Eq.~\eqref{eq:hansen_coeff}. This allows us to write $\mathcal{A}_j$ as a Fourier series in the mean longitudes $\lambda_i$, from which we extract its secular part $\mathcal{A}_j^{(0,0)}$ by isolating the terms with zero frequency in $\lambda_i$.
 The secular component of the direct part of the perturbing function at order $k$ is then given by
\begin{equation}
\label{eq:secular_directpart_orderk}
\langle U_1 \rangle = -\frac{G m_1 m_2}{a_2} \sum_{j=0}^k \mathcal{A}_j^{(0,0)}.
\end{equation}
The derivation of $\mathcal{A}_2$ and its secular component $\mathcal{A}_2^{(0,0)}$ is provided in Appendix~\ref{appendix:LL_order2} as an additional example of the steps outlined here.

While the algebraic complexity increases with the order $k$, the structure of the expansion remains recursive and systematic. This makes the approach well-suited for symbolic computation using a computer algebra system. We implement this formalism using \textsc{TRIP}, a computer algebra system designed for handling perturbation series in celestial mechanics \citep{Gastineau2011,TRIP}. It includes built-in libraries for manipulating both Hansen and Laplace coefficients, and evaluating them when needed for numerical application.

\section{Convergence of the Laplace-Legendre expansion}
\label{sec:numerical_convergence}
The convergence of truncated secular expansions of the disturbing function has been investigated in several previous studies, primarily through numerical assessments of the accuracy of high-order expansions. In \citep{Laskar1985}, the convergence of the Laplace expansion at second order of the masses is analyzed for all planetary pairs in the Solar System with respect to the truncation in the harmonics of the first order expansion. For the coplanar 3-body problem, Libert \& Henrard \citeyearpar{Libert2005} considered Hamiltonian expansions carried to degree 12 in eccentricity and showed that the resulting secular models accurately reproduce the long-term dynamics of exoplanetary systems with large eccentricities, while also demonstrating that the series of secular terms converges better than the series associated with the full perturbation. Migaszewski \& Goździewski \citeyearpar{Migaszewski2008} performed a similar convergence study of the Legendre expansion, extending it up to order 24 in the semi-major axis ratio and applying it to a range of exoplanetary configurations. They showed that the expansion converges rapidly for sufficiently hierarchical systems, while the convergence becomes progressively slower as the system becomes more compact.

In this section, we compare the convergence properties of the Laplace–Legendre expansion introduced in Sect.~\ref{sec:LL_formulation} with those of the classical Laplace and Legendre expansions. In line with the aforementioned studies, we focus here on a numerical assessment of convergence based on the evaluation of the series at different truncation orders. In the celebrated  chapter VIII on Formal calculus, in {\it Les Méthodes Nouvelles de la Mécanique Céleste}, Poincaré \citeyearpar{Poincare1892} distinguishes the convergence in the sens of the astronomers and mathematicians, and reminds that both approaches are legitimate. 
As discussed in Sect.~\ref{sec:Laplace_expansion}, we know that for sufficiently small values of the ratio of semi major axis, $\alpha$, eccentricities and inclinations, the series expansion \eqref{sec:Laplace_expansion} is convergent. But in the present work, we consider formal expansions and address the convergence questions in a numerical way, in the sense of the astronomers, without addressing more mathematical questions such as the domain of convergence of the series, the size of the remainder, or the decay rate of the terms. 
Our goal is to assess the accuracy of each series in approximating the secular direct part of the perturbing function, $\langle U_1 \rangle$, as a function of the truncation order. As in the previous section, we focus on the planar problem, where $\langle U_1 \rangle$ depends on four parameters: the eccentricities $e_i$, the difference of arguments of pericenter $\Delta\omega = \omega_1 - \omega_2$, and the semi-major axis ratio $\alpha $. For simplicity, we assume equal eccentricities, $e_1=e_2=e$, and explore a range of values in $e$ and $\alpha$, from nearly circular ($e \ll 1$) to moderately eccentric orbits ($e \sim 0.2$), and from hierarchical ($\alpha \ll 1$) to moderately compact ($\alpha \sim 0.2$) configurations. We also consider two representative values of $\Delta\omega$, namely $0^\circ$ and $180^\circ$. 

For each set of parameters, we compute a reference value of $\langle U_1 \rangle$ by numerically averaging the inverse of mutual distance $a_2/\Delta$ over the mean anomalies. In practice, we perform the average over the eccentric anomalies $E_i$ instead, using the standard relation $\mathrm{d}M_i = (1 - e_i\cos E_i) \mathrm{d}E_i$. We discretize the integral on a uniform 2D grid in $E_i$, evaluate $a_2/\Delta$ at each grid point, and compute the average over the grid using the trapezoidal rule. We denote this reference value as $\langle U_1 \rangle_{\text{num}}$. We then evaluate $\langle U_1 \rangle$ using the three expansions at increasing truncation orders $N$, defined as follows. For the Legendre expansion, $N$ denotes the truncation order in the semi-major axis ratio $\alpha$ used in Eq.~\eqref{eq:Legendre_secular} to compute the secular part. For the Laplace expansion, $N$ represents the maximum total degree in eccentricities, and the secular part is obtained by truncating Eq.~\eqref{eq:Laplace_planar} to retain all terms such that $n_1 + \bar n_1 +n_2 + \bar n_2 \leq N$.
In the Laplace–Legendre formulation, $N$ refers to the upper index in the sum $\sum_{j=0}^{N} \mathcal{A}_j^{(0,0)}$, so that $N+1$ terms are retained from the Laplace expansion (Eq.~\eqref{eq:distance_Laplace}). The secular contribution at each order is computed using Eq.~\eqref{eq:secular_directpart_orderk}, with each term evaluated according to the formalism presented in Sect.~\ref{sec:LL_formulation}. It should be noted that for each $k>0$, the expression for the term $\mathcal{A}_k^{(0,0)}$ involves an infinite summation over Laplace coefficients $b_{k+\frac{1}{2}}^{(l)}$; see Eq.~\eqref{eq:A1_Laplace_secular} for $\mathcal{A}_1^{(0,0)}$ as an example. These coefficients decay exponentially as $l \to \infty$, since they are Fourier coefficients of an analytic function when $\alpha < 1$. This allows the summation to be truncated for numerical implementation. One option is to define a global truncation limit $L$ that ensures a given precision across all terms considered. Alternatively, an adaptive approach consists in selecting, for each $k$, a value of $L$ large enough to ensure $\mathcal{A}_k^{(0,0)}$ converges to the requested precision. In our study, convergence of each $\mathcal{A}_k^{(0,0)}$ was verified to machine precision ($\sim10^{-16}$) over all parameter regimes explored in the present section. This includes configurations with
$e,\alpha\in [0,0.2]$, as well as two limiting cases: nearly circular but compact configurations with $e=0.01$ and
$\alpha\in\{0.4,0.5\}$, and hierarchical, highly eccentric configurations with $e=0.6$ and $\alpha\in\{0.001,0.01\}$. The truncation limits $L_k$ adopted for each $\mathcal{A}_k^{(0,0)}$, together with the corresponding number of terms in its resulting expression, are reported in Appendix~\ref{appendix:truncation_terms} for orbital configurations with $e,\alpha\in [0,0.2]$. For completeness, we also report the number of terms obtained at a given order in the classical Laplace and Legendre expansions. In all cases, the relative error is defined as 
\begin{equation}
\label{eq:error}
\text{relative error} = \left \vert 1-\frac{\langle U_1 \rangle}{\langle U_1\rangle_{\text{num} }}\right \vert.
\end{equation}
\begin{figure}[htp!]
  \centering
  \includegraphics[width=\textwidth]{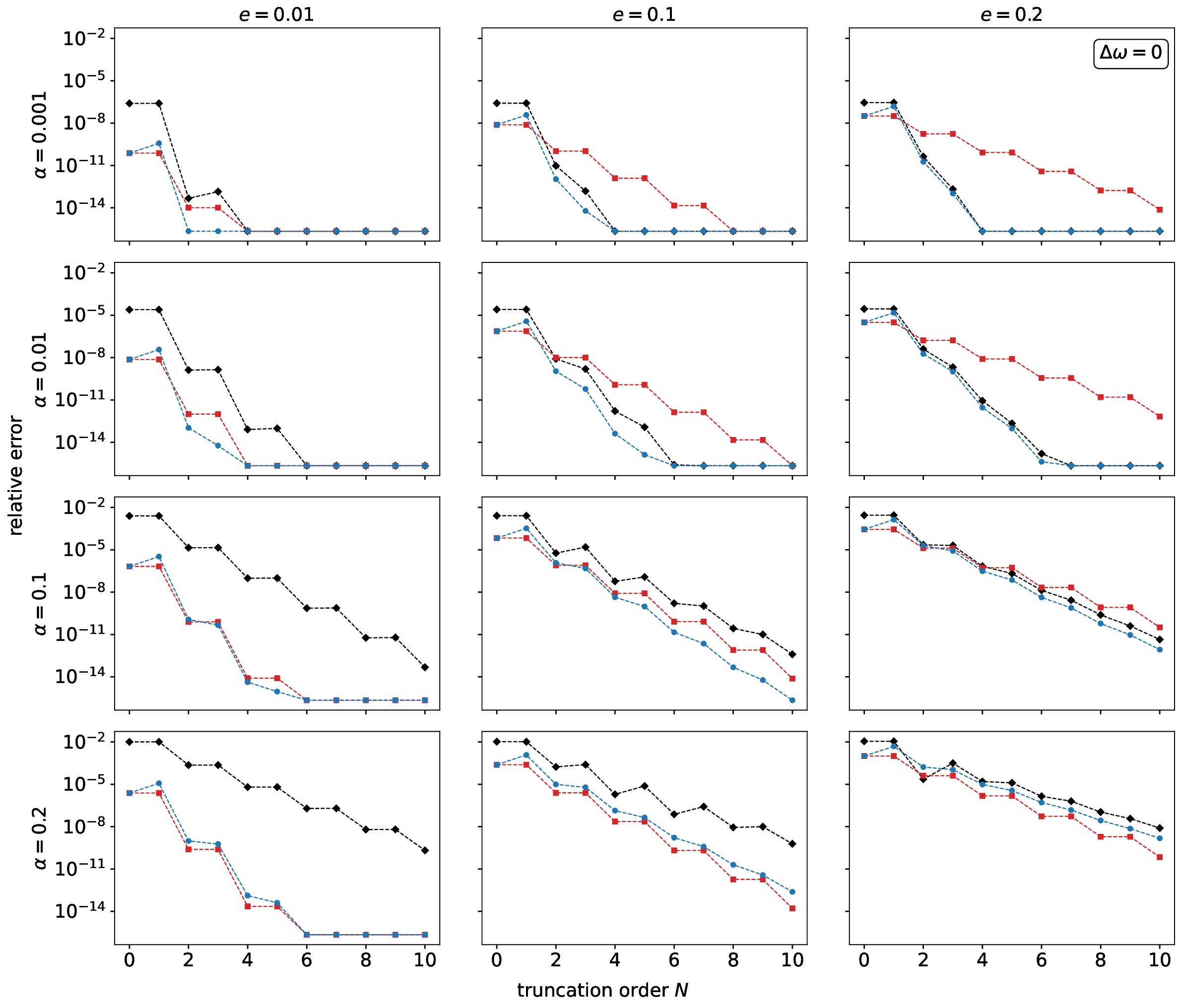}
  \vspace{-0.5 cm}
  \includegraphics[width=\textwidth]{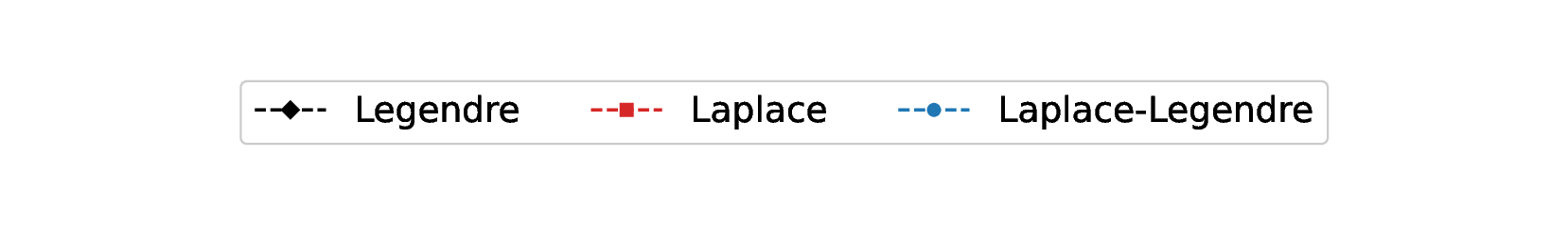}  
  \caption{
     Relative error in the computation of $\langle U_1 \rangle$ as a function of the truncation order $N$, comparing the Legendre (black), Laplace (red), and Laplace–Legendre (blue) expansions. Each panel corresponds to a specific configuration of eccentricity and semi-major axis ratio. Columns show increasing eccentricities $e = 0.01$, $0.1$, and $0.2$ (left to right), while rows correspond to increasing semi-major axis ratios $\alpha = 0.001$, $0.01$, $0.1$, and $0.2$ (top to bottom). The value of $\Delta \omega$ is fixed at $0^\circ$. All the numerical evaluations are performed to double precision.
  }
  \label{fig:convergence_grid_phi0}
\end{figure}

\begin{figure}[htp!]
  \centering
  \includegraphics[width=\textwidth]{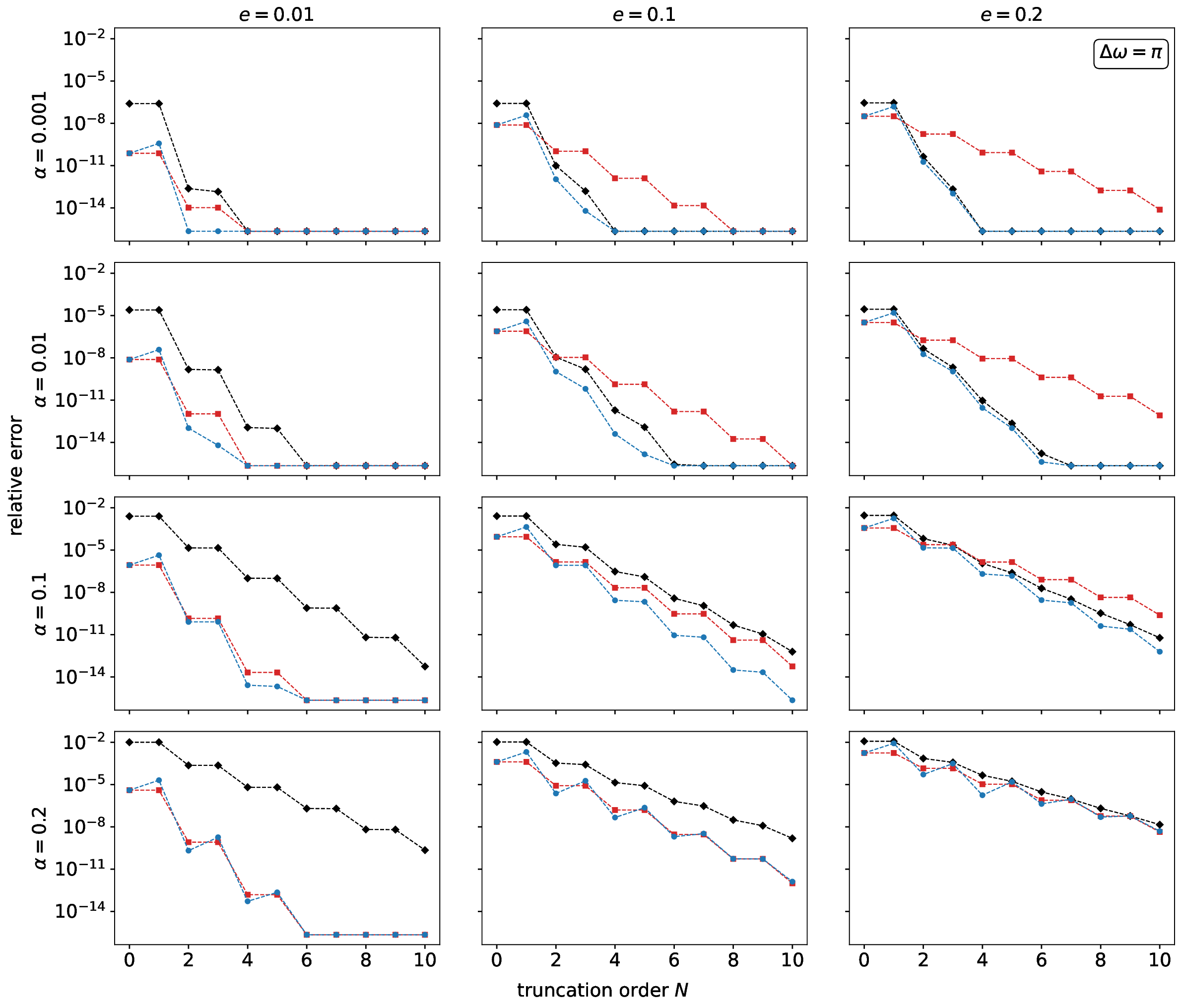}
  \vspace{-0.5 cm}
  \includegraphics[width=\textwidth]{figures/full_legend_Aorder.eps}  
  \caption{
     Same as Fig.~\ref{fig:convergence_grid_phi0}, but for a fixed value of $\Delta \omega = 180^\circ$.
  }
  \label{fig:convergence_grid_phi180}
\end{figure}

Figure~\ref{fig:convergence_grid_phi0} displays the relative error in computing $\langle U_1 \rangle$ as a function of the truncation order $N$ for the three expansions, with $\Delta\omega = 0^\circ$. The grid layout allows for a direct comparison across different dynamical regimes: each row corresponds to a fixed value of $\alpha$, and each column to a fixed value of the eccentricities (with $e_1 = e_2$). 
The Laplace–Legendre expansion improves over the Legendre expansion, in many of the tested configurations. This improvement is most pronounced in the low-eccentricity regime (e.g., for $e = 0.01$), where it reaches high accuracy with fewer orders, particularly as $\alpha$ increases. This is because it is constructed as a series in $V$, which scales as $\alpha e$ (see Eq.~\eqref{eq:V_small_e}), whereas the Legendre expansion (Eq.~\eqref{eq:Legendre_secular}) is a series in $\alpha$, and does not benefit from the smallness of $e$. In this regime, its performance closely matches that of the Laplace expansion, which is already well suited for nearly circular orbits, independently of $\alpha$. An intermediate regime ($e = 0.1$–$0.2$, $\alpha = 0.1$) appears where the Laplace–Legendre expansion can outperform both classical methods provided that the truncation order $N$ is sufficiently high. At higher eccentricities ($e = 0.2$), the Laplace–Legendre expansion behaves similarly to the Legendre expansion and shows moderate improvements across all values of $\alpha$. In this regime, both methods remain more accurate than the Laplace expansion in the hierarchical limit ($\alpha = 0.001$, $0.01$), whereas the Laplace expansion provides the best accuracy in compact configurations ($\alpha = 0.2$).

Figure~\ref{fig:convergence_grid_phi180} presents the same comparison for $\Delta\omega = 180^\circ$. The qualitative behavior of the three expansions remains consistent with the $\Delta\omega = 0^\circ$, particularly in the small $e$ and $\alpha$ regimes. A notable difference arises in the moderately eccentric and compact configurations ($e = 0.1$–$0.2$, $\alpha = 0.2$), where the Laplace–Legendre expansion shows improved convergence and matches the performance of the Laplace expansion more closely. We also note that Eq.~\eqref{eq:distance_Laplace} is an alternating series when $V>0$. This can lead to an non-monotonic behavior in the convergence of the Laplace-Legendre expansion, where the relative error may increase at certain odd truncation orders before improving again.
\begin{figure}[htp!]
  \centering
  \includegraphics[width=\textwidth]{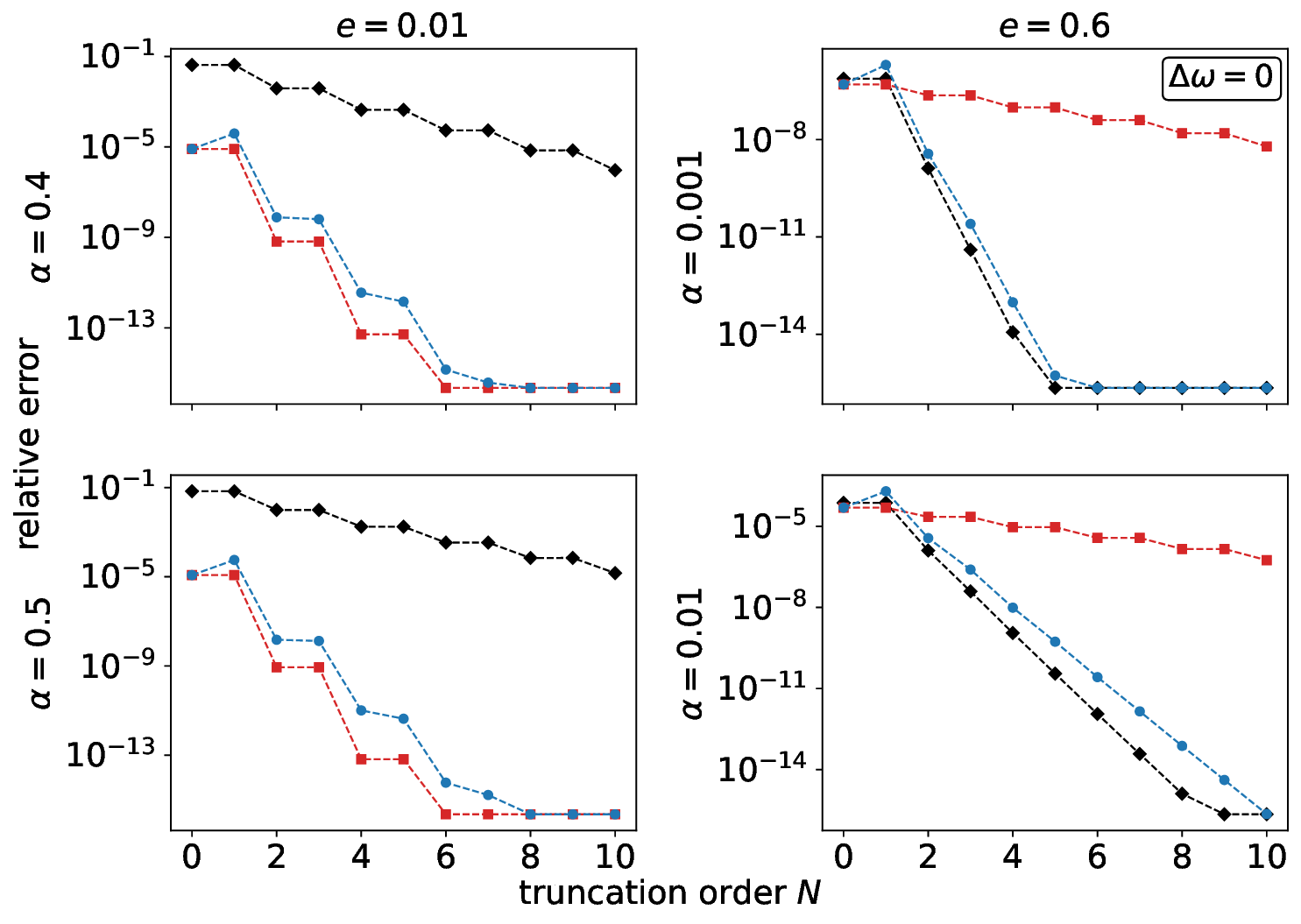}
  \vspace{-0.5 cm}
  \includegraphics[width=\textwidth]{figures/full_legend_Aorder.eps}  
  \caption{
     Relative error in the computation of $\langle U_1 \rangle$ as a function of the truncation order $N$, comparing the Legendre (black), Laplace (red), and Laplace–Legendre (blue) expansions, for $\Delta \omega = 0^\circ$. The left column corresponds to the nearly circular compact regime ($e = 0.01$, $\alpha = 0.4, 0.5$), while the right column shows the hierarchical eccentric one ($e = 0.6$, $\alpha = 0.001, 0.01$).}
  \label{fig:convergence_grid_limits}
\end{figure}

We further explore the performance of the Laplace–Legendre expansion in two contrasting regimes: the nearly circular, compact configuration ($e = 0.01$, $\alpha = 0.4,0.5$), where the Laplace expansion is known to perform best, and the hierarchical, highly eccentric case ($e = 0.6$, $\alpha \ll 1$), where the Legendre expansion yields the most accurate results. These regimes are illustrated in Fig.~\ref{fig:convergence_grid_limits}, which shows the relative error as a function of truncation order $N$ for the three expansions. In both regimes, the Laplace–Legendre expansion tracks the classical expansion that is best suited to the respective regime. The slightly reduced accuracy seen in Fig.~\ref{fig:convergence_grid_limits} could be justified by the fact that the new formulation orders the series in powers of the parameter $V$, rather than in the small parameter that is appropriate to each regime ($e$ for Laplace, $\alpha$ for Legendre). As a result, in these limiting cases the ordering is not as locally optimal as in the corresponding classical expansion. This is a trade-off for the optimality of the new expansion in the regime $e,\alpha \to 0$. Anyway, the clear advantage of the new expansion is that now we have a global formulation that bridges the Laplace and Legendre regimes. This broader applicability comes at the cost of a larger number of terms retained at a given expansion order compared to the classical Laplace and Legendre expansions (see Appendix~\ref{appendix:truncation_terms}).

Finally, we note that when both $e$ and $\alpha$ are not small (e.g, $e,\alpha > 0.2$), the convergence of the Laplace-Legendre expansion deteriorates and becomes progressively slower, similarly to the classical expansions. This behavior is expected, as in such regimes the parameter $V$ is no longer small and the assumptions underlying the present expansion are no longer satisfied (see Sect.~\ref{sec:LL_formulation}).

\section{Conclusion}\label{sec:Conclusion}
In this work, we explored an alternative expansion of the disturbing function for the planar planetary 3-body problem. This formulation, referred to as the Laplace–Legendre expansion, combines elements from both the classical Laplace and Legendre series. It uses the same series structure as the Laplace formulation, but differs in that each term of the expansion retains its full dependence in both the semi-major axis ratio and eccentricity. The method provides, for example, an alternative means to construct the secular Hamiltonian suitable for modeling the long-term evolution of planetary systems.

We evaluated the convergence of this expansion across several choices of eccentricities $e$ and semi-major axis ratios $\alpha$, and compared its performance with that of the Laplace and Legendre series. We explored configurations characterized by $e,\alpha\in[0,0.2]$, as well as two limiting regimes: nearly circular, compact systems with $e=0.01$ and $\alpha\in\{0.4,0.5\}$, and hierarchical, highly eccentric systems with $e=0.6$ and $\alpha\in\{0.001,0.01\}$. The results show that the Laplace–Legendre expansion performs best in regimes where both $e$ and $\alpha$ are small (e.g., $e = 0.01$, $\alpha = 0.01$). It can also outperform both classical methods in an intermediate regime, where neither parameter is particularly small but their product remains small (e.g., $e = 0.1$, $\alpha = 0.1$). In the hierarchical limit $(\alpha < 0.1)$, the Laplace–Legendre expansion improves upon the Legendre expansion at low eccentricities ($e<0.1$), while at higher eccentricities ($e\geq 0.2$) it closely reproduces the performance of the Legendre expansion. In the nearly circular regime $(e \lesssim 0.01)$, and as long as $\alpha$ remains moderate, its behavior closely matches that of the Laplace expansion. In the last two regimes, combining the Laplace and Legendre approaches does not necessarily lead to an expansion that outperforms both classical methods at the same time. Instead, the Laplace–Legendre expansion tends to match the method that is best suited to each limit. In the extremes of these limits (namely, when $\alpha \ll 1$ and $e$ is large, or when $e \sim 0$ while $\alpha$ is large), the new expansion may exhibit slightly reduced accuracy compared to the classical method most appropriate for that regime. Nevertheless, its clear advantage is to span in a single formulation both the Laplace and Legendre expansions. It may therefore be useful when the system transitions between different dynamical regimes, or when the prevailing regime cannot be determined a priori.

Overall, the Laplace–Legendre expansion provides consistent performance across a wide range of orbital parameters which includes the domain of applicability of both the Laplace and Legendre expansions, making it an interesting option for secular studies. The price to pay for this broader applicability is an increasing complexity of the analytical expression of the Hamiltonian at each order with respect to the classical expansions, which results in a larger number of terms involved at a given order. In particular, the expression of the secular Hamiltonian involves an infinite summation over the Laplace coefficients and therefore is not in a closed form. This naturally calls for the use of a computer algebra system to manipulate large expressions, particularly for the spatial problem, which we did not address in this exploratory work. Extending the Laplace–Legendre formulation to the spatial case is feasible along the same lines of this work, but would substantially increase the algebraic complexity, as is already apparent in the classical expansions when moving from the planar to the spatial formulation \citep[e.g.,][]{Laskar2010}. While a dedicated study would be required to quantify convergence in the spatial case, the new formulation remains a formal expansion in the parameter $V$. We therefore expect similar qualitative convergence behavior in the regime where $V$ remains small, i.e., for small to moderate eccentricities, inclinations and semi-major axis ratios.

\begin{appendices}
\section{Laplace-Legendre expansion of order 2}
\label{appendix:LL_order2}
We present here the Laplace–Legendre expansion of order $2$, defined by
\begin{equation}
\label{eq:LL_order2_def}
\frac{a_2}{\Delta} = \mathcal{A}_0 + \mathcal{A}_1 + \mathcal{A}_2,
\end{equation}
where $\mathcal{A}_0$ and $\mathcal{A}_1$ are the first and second terms of the Laplace expansion, given in Sect.~\ref{sec:LL_formulation}, and $\mathcal{A}_2$ reads
\begin{equation}
\label{eq:A2_expr}
\mathcal{A}_2 = \frac{3}{8} \frac{a_2}{r_2} V^2 A^{-5/2}.
\end{equation}
As in previous orders, we first express $A^{-5/2}$ using Laplace coefficients as
\begin{equation}
A^{-5/2} = \frac{1}{2} \sum_{l \in \mathbb{Z}} b_{5/2}^{(l)}(\alpha)  \E^{\iota l (\lambda_1 - \lambda_2)}.
\end{equation}
We then expand $\dfrac{a_2}{r_2} V^2 $ as a Fourier series in $\lambda_i$, using Hansen coefficients (see Eq.~\ref{eq:hansen_coeff}),
\begin{equation}
\label{eq:V^2_lambda}
\begin{aligned}
\frac{a_2}{r_2} V^2 =& \, 2 \alpha ^2\sum_{k_2 \in \mathbb{Z}} X_{k_2}^{-1,0} \, \E^{\iota k_2(\lambda_2 - \omega_2)} \left(  (1+ \frac{\alpha^2}{2}) - 2 \alpha  \cos(\lambda_1 - \lambda_2) + \cos{2(\lambda_1 - \lambda_2)}  \right) \\
& +2 \vphantom{\biggl(\biggr)}\alpha^2 \sum_{k_1, k_2 \in \mathbb{Z}} \E^{\iota[k_1(\lambda_1 - \omega_1) + k_2(\lambda_2 - \omega_2)]}  \\
&\qquad  \qquad \times\Biggl[ \left(1-\alpha^2 + 2\alpha \cos(\lambda_1 - \lambda_2)\right)  X_{k_1}^{2,0} X_{k_2}^{-3,0} + \frac{\alpha^2}{2} X_{k_1}^{4,0} X_{k_2}^{-5,0} \\
& \qquad  \qquad + \biggl( \left(\alpha- 2 \cos(\lambda_1 - \lambda_2)\right) X_{k_1}^{1,1} X_{k_2}^{-2,-1} - \alpha X_{k_1}^{3,1} X_{k_2}^{-4,-1}\biggr)\E^{\iota(\omega_1 - \omega_2)} \\
& \qquad  \qquad + \biggl( \left(\alpha- 2 \cos(\lambda_1 - \lambda_2)\right) X_{k_1}^{1,-1} X_{k_2}^{-2,1} - \alpha X_{k_1}^{3,-1} X_{k_2}^{-4,1} \biggr)\E^{-\iota(\omega_1 - \omega_2)} \\
& \vphantom{\biggl(\biggr)}\qquad  \qquad  + \frac{1}{2} X_{k_1}^{2,2} X_{k_2}^{-3,-2}  \E^{2\iota(\omega_1 - \omega_2)}  + \frac{1}{2} X_{k_1}^{2,-2} X_{k_2}^{-3,2}   \E^{-2\iota(\omega_1 - \omega_2)}\Biggr],
\end{aligned}
\end{equation} 
Substituting this expression together with the expansion of $A^{-5/2}$ into the definition of $\mathcal{A}_2$, we get
\begingroup
\allowdisplaybreaks
\begin{align}
\label{eq:A2_Laplace}
\mathcal{A}_2 =& \, \frac{3}{8} \alpha ^2\sum_{l,k_2 \in \mathbb{Z}}  b_{5/2}^{(l)}(\alpha) \, X_{k_2}^{-1,0} \E^{\iota k_2(\lambda_2 - \omega_2)}  \E^{\iota l (\lambda_1 - \lambda_2)} \notag  \\
& \qquad \qquad \times \left(  (1+ \frac{\alpha^2}{2}) - 2 \alpha  \cos(\lambda_1 - \lambda_2) + \cos{2(\lambda_1 - \lambda_2)}  \right) \notag \\
& +\frac{3}{8}  \vphantom{\biggl(\biggr)}\alpha^2 \sum_{l,k_1, k_2 \in \mathbb{Z}} b_{5/2}^{(l)}(\alpha)   \E^{\iota[(k_1 + l)\lambda_1 + (k_2 - l)\lambda_2]} \, \E^{-\iota(k_1 \omega_1 + k_2 \omega_2)} \notag \\
&\qquad \qquad \times\Biggl[ \left(1-\alpha^2 + 2\alpha \cos(\lambda_1 - \lambda_2)\right)  X_{k_1}^{2,0} X_{k_2}^{-3,0} + \frac{\alpha^2}{2} X_{k_1}^{4,0} X_{k_2}^{-5,0} \notag \\ 
& \qquad \qquad  + \biggl( \left(\alpha- 2 \cos(\lambda_1 - \lambda_2)\right) X_{k_1}^{1,1} X_{k_2}^{-2,-1} - \alpha X_{k_1}^{3,1} X_{k_2}^{-4,-1}\biggr)\E^{\iota(\omega_1 - \omega_2)} \notag \\
& \qquad \qquad  + \biggl( \left(\alpha- 2 \cos(\lambda_1 - \lambda_2)\right) X_{k_1}^{1,-1} X_{k_2}^{-2,1} - \alpha X_{k_1}^{3,-1} X_{k_2}^{-4,1} \biggr)\E^{-\iota(\omega_1 - \omega_2)} \notag \\
&\qquad \qquad  \vphantom{\biggl(\biggr)}+ \frac{1}{2} X_{k_1}^{2,2} X_{k_2}^{-3,-2}  \E^{2\iota(\omega_1 - \omega_2)} + \frac{1}{2} X_{k_1}^{2,-2} X_{k_2}^{-3,2}   \E^{-2\iota(\omega_1 - \omega_2)}\Biggr].
\end{align}    
\endgroup
To extract the secular part of $\mathcal{A}_2$, we retain only the terms with zero frequency in the mean longitudes. In the first sum, this corresponds to $k_2 = 0$ and $l \in \{ -2,-1, 0, 1, 2 \}$. In the second sum, secular contributions arise from specific index combinations related to the angular dependence of each term. In particular, for each $l \in \mathbb{Z}$ we retain the couples $(k_1,k_2)\in \bigl\{(-l-1,l+1),(-l,l),(-l+1,l-1)\bigr \}$. This yields

\begin{equation}
\label{eq:A2_Laplace_secular}
\begin{aligned}
\mathcal{A}_{2}^{(0,0)} &= \frac{3}{8} \alpha^2 \left(  (1+ \frac{\alpha^2}{2}) b_{5/2}^{(0)} (\alpha)- 2 \alpha b_{5/2}^{(1)} (\alpha)+ b_{5/2}^{(2)}(\alpha) \right)\\[1ex]
&+\frac{3}{8} \alpha^2  \sum_{l \in \mathbb{Z}} b_{5/2}^{(l)}  \Bigg[ \Bigg( (1-\alpha^2) X_{-l}^{2,0}(e_1) \,  X_l^{-3,0}(e_2) +\frac{\alpha^2}{2} X_{-l}^{4,0}(e_1) \,  X_l^{-5,0}(e_2) \\[1ex]
& -  X_{-l+1}^{1,1}(e_1) \, X_{l-1}^{-2,-1}(e_2)-  X_{-l-1}^{1,-1}(e_1) \, X_{l+1}^{-2,1}(e_2) \Bigg) \E^{\iota l(\omega_1 - \omega_2)}\\
 &  + \alpha \Bigg(  X_{-l-1}^{2,0}(e_1) \, X_{l+1}^{-3,0}(e_2) - X_{-l}^{3,1}(e_1)\, X_l^{-4,-1}(e_2)
+ X_{-l}^{1,1}(e_1) \,X_l^{-2,-1}(e_2) \Bigg)\E^{\iota (l+1)(\omega_1 - \omega_2)}\\
 & + \alpha \Bigg( X_{-l+1}^{2,0}(e_1) \, X_{l-1}^{-3,0}(e_2)
- X_{-l}^{3,-1}(e_1) \, X_l^{-4,1}(e_2) + X_{-l}^{1,-1}(e_1) \, X_l^{-2,1}(e_2) \Bigg)\E^{\iota (l-1)(\omega_1 - \omega_2)}\\
\end{aligned}
\end{equation}   
\begin{equation*}
\begin{aligned}
  &  +\Bigg(  - X_{-l-1}^{1,1}(e_1) \, X_{l+1}^{-2,-1}(e_2)
+\frac{1}{2} X_{-l}^{2,2}(e_1) \, X_l^{-3,-2}(e_2) \Bigg)\E^{\iota (l+2)(\omega_1 - \omega_2)}\\
 &  +\Bigg( -X_{-l+1}^{1,-1}(e_1) \, X_{l-1}^{-2,1}(e_2)
+ \frac{1}{2}X_{-l}^{2,-2}(e_1) \, X_l^{-3,2}(e_2) \Bigg)\E^{\iota (l-2)(\omega_1 - \omega_2)}
 \Bigg]. 
\end{aligned}
\end{equation*}

The secular component of direct part of the perturbing function $U_1$ is then given by
\begin{equation}
\langle U_1 \rangle = -\frac{G m_1 m_2}{a_2} \left( \mathcal{A}_0^{(0,0)} + \mathcal{A}_1^{(0,0)} + \mathcal{A}_2^{(0,0)} \right).
\end{equation}

\section{Number of terms in the secular expansions}
\label{appendix:truncation_terms}
This appendix reports the number of terms involved in the three secular expansions considered in this work.

For the classical Laplace and Legendre expansions, Table~\ref{tab:number_of_terms_classical} reports the number of terms appearing at a given expansion order in the secular direct part of the disturbing function, that is, $\langle U_1\rangle$. In other words, the table reports the number of terms contributing to the first-order secular Hamiltonian. In the Laplace formulation, this corresponds to the coefficient $C_{0,0}$ in Eq.~\eqref{eq:Laplace_planar}, while in the Legendre formulation it corresponds to the secular expansion given by Eq.~\eqref{eq:Legendre_secular}.

\begin{table*}[htp!]
\centering
\renewcommand{\arraystretch}{0.9}
\begin{tabular}{c|c|c}
\hline\hline
Expansion order & Laplace expansion & Legendre expansion \\
\hline
$0$  & $1$   & $1$ \\
$2$  & $8$   & $1$ \\
$4$  & $25$  & $3$ \\
$6$  & $70$  & $5$ \\
$8$  & $153$ & $7$ \\
$10$ & $294$ & $9$ \\
\hline
\end{tabular}
\vspace*{1mm}
\caption{Number of terms appearing at a given order in the secular direct part of disturbing function $\langle U_1 \rangle$, for the classical Laplace ($C_{0,0}$ in Eq.~\ref{eq:Laplace_planar}) and Legendre (Eq.~\ref{eq:Legendre_secular}) expansions.}
\label{tab:number_of_terms_classical}
\end{table*}
In the Laplace–Legendre expansion, for a given $k>0$, the expression for the term $\mathcal{A}_k^{(0,0)}$ involves an infinite summation over Laplace coefficients $b_{k+\frac{1}{2}}^{(l)}$ that must be truncated for numerical evaluation (see, e.g., Eqs.~\eqref{eq:A1_Laplace_secular} and~\eqref{eq:A2_Laplace_secular} for $k=1,2$). Tables~\ref{tab:truncation_grid} and~\ref{tab:truncation_grid_single} report the truncation limits $L_k$ adopted to ensure double-precision ($\sim10^{-16}$) and single-precision ($\sim10^{-8}$) accuracy, respectively, in the numerical evaluation of $\mathcal{A}_k^{(0,0)}$, together with the corresponding number of terms in its resulting expression. These values are reported for the orbital configurations considered in Sect.~4, in particular, for $e,\alpha \in [0,0.2]$ and $\Delta\omega =0$.
\begin{table*}[t]
\centering
\renewcommand{\arraystretch}{1}
\begin{adjustbox}{max width=\linewidth}
\begin{tabular}{c|c| c| c}
\hline
\hline
\diagbox{$\alpha$}{$e$} & $0.01$ & $0.10$ & $0.20$ \\
\hline
$0.001$ &
\celltab{1&3&20 \\2 & 3& 89} &
\celltab{1 & 4 &26 \\ 2&4&119\\3&5&415\\4&6&1153} &
\celltab{1&4&26\\2&5&149\\3&6&503\\4&6&1153} \\
\hline
$0.01$ &
\celltab{1&4&26 \\2 & 5& 149\\ 3&6&503\\4&7&1363} &
\celltab{1 & 5 &32 \\ 2&5&149\\3&6&503\\4&7&1363 \\5&8&2933\\6&9&6099} &
\celltab{1&6&38\\2&6&179\\3&7&591\\4&7&1363\\5&9&3335 \\6&9&6099} \\
\hline
$0.1$ &
\celltab{1&5&32\\2&6&179\\3&7&591\\4&8&1573\\5&10&3737 \\6&11&7611}  &
\celltab{1&7&44\\2&8&239\\3&9&767\\4&10&1993\\5&12&4541 \\6&13&9123\\7&14&15761\\8&14&24648\\9&17&44569\\10&18&69753}  &
\celltab{1&8&50\\2&9&269\\3&11&943\\4&12&2413\\5&13&4943 \\6&14&9879\\7&16&18209\\8&17&30480\\9&18&47447\\10&19&74043}  \\
\hline
$0.2$ &
\celltab{1&5&32\\2&6&179\\3&8&679\\4&8&1573\\5&11&4139 \\6&11&7611}  &
\celltab{1&7&44\\2&9&269\\3&11&943\\4&12&2143\\5&14&5345 \\6&15&10635\\7&16&18209\\8&18&32424\\9&19&50325\\10& 21 &82623}  &
\celltab{1&9&55\\2&11&329\\3&13&1119\\4&14&2833\\5&16&6149 \\6&16&11391\\7&19&21881\\8&20&36312\\9&22&58959\\10& 23& 91203}  \\
\hline
\end{tabular}
\end{adjustbox}
\vspace*{1mm}
\caption{\textbf{Truncation limits and number of terms involved in the numerical evaluation of the Laplace–Legendre expansion.} For each pair $(e,\alpha)$ and each expansion order $k$, we report the truncation limit $L_k$ needed to ensure double-precision accuracy ($\sim10^{-16}$) in the evaluation of $\mathcal{A}_k^{(0,0)}$ (see, e.g., Eqs.~\eqref{eq:A1_Laplace_secular} and~\eqref{eq:A2_Laplace_secular} for $k=1,2$), together with the corresponding number of terms in its resulting expression. Results are reported up to the lowest expansion order $k$ for which $\langle U_1 \rangle$ reaches double-precision agreement with the numerical average $\langle U_1 \rangle_{\text{num}}$; when such agreement is not achieved, results are shown up to the highest expansion order considered in this work $(k=10)$.}
\label{tab:truncation_grid}
\end{table*}

\begin{table*}[t]
\centering
\renewcommand{\arraystretch}{1.1}
\begin{adjustbox}{max width=\linewidth}
\begin{tabular}{c|c| c| c}
\hline
\hline
\diagbox{$\alpha$}{$e$} & $0.01$ & $0.10$ & $0.20$ \\
\hline
$0.001$ &
\celltab{1&2&14 \\2 & 2& 59} &
\celltab{1 & 2 &14 \\ 2&3&89\\3&3&239\\4&4&733} &
\celltab{1&3&20\\2&3&89\\3&4&327\\4&4&733} \\
\hline
$0.01$ &
\celltab{1&2&14 \\2 & 3& 89\\ 3&4&327\\4&5&943} &
\celltab{1 & 3 &20 \\ 2&3&89\\3&4&327\\4&5&943 \\5&6&2129\\6&6&3831} &
\celltab{1&3&20\\2&3&89\\3&5&415\\4&5&943\\5&6&2129 \\6&7&4587} \\
\hline
$0.1$ &
\celltab{1&3&20\\2&4&119\\3&5&415\\4&6&1153\\5&8&2933 \\6&8&5343}  &
\celltab{1&4&26\\2&5&149\\3&6&503\\4&7&1363\\5&8&2933 \\6&9&6099\\7&11&12089\\8&12&20760\\9& 12&30179\\10& 13 &48303} 
&
\celltab{1&5&32\\2&5&149\\3&7&591\\4&8&1573\\5&9&3335 \\6&10&6855\\7&11&12089\\8&12&20760\\9&12&30179\\10& 13 & 48303}   \\
\hline
$0.2$ &
\celltab{1&3&20\\2&4&119\\3&6&503\\4&7&1363\\5&9&3335 \\6&10&6855}  &
\celltab{1&4&26\\2&5&149\\3&7&594\\4&8&1573\\5&10&3737 \\6&11&7611\\7& 13&14537 \\8& 14& 24648\\9& 15&38813 \\10& 17  & 65463 }  &
\celltab{1&5&32\\2&6&179\\3&8&679\\4&9&1783\\5& 11&4139  \\6& 12&8367\\7& 14 & 15761\\8& 15&26592\\9&16 &41691\\10& 17 & 65463 }  \\
\hline
\end{tabular}
\end{adjustbox}
\vspace*{2mm}
\caption{\footnotesize Same as Table~\ref{tab:truncation_grid}, but the truncation limit $L_k$ is chosen to ensure single-precision accuracy $\sim 10^{-8}$ in computing $\mathcal{A}_k^{(0,0)}$.}
\label{tab:truncation_grid_single}
\end{table*}

The tables show that the Laplace–Legendre expansion generally involves a larger number of terms than the classical Laplace and Legendre expansions, even at low expansion orders and for moderate values of $(e,\alpha)$, and increasingly so at higher orders. This is related to the structure of the Laplace–Legendre formulation, in which the dependence on eccentricity and semi-major axis ratio is treated exactly at each order, leading to coefficients $\mathcal{A}_k^{(0,0)}$ with a more complex structure. As a result, the number of retained terms depends on the truncation strategy and precision requirement adopted in the numerical evaluation.

\end{appendices}

\clearpage





\bmhead{Acknowledgements}
This work has been supported by the French Agence Nationale de la Recherche (AstroMeso ANR-19-CE31-0002-01) and the European Research Council (ERC) under the European Union’s Horizon 2020 research and innovation program (Advanced Grant AstroGeo-885250).

\bibliography{sn-bibliography}

\end{document}